\documentstyle{ar209}

\def\be{\begin{equation}}
\def\ee{\end{equation}}
\def\bea{\begin{eqnarray}}
\def\eea{\end{eqnarray}}

\def\lsim{\mbox{{\scriptsize \raisebox{-.9ex}
      {$\;\stackrel{{\textstyle <}}{\sim}\,$} }} }

\def\eg{{\it e.g.\ }}
\def\etal{{\it et al.\ }}
\newcommand{\e}{{\mbox{e}}}

\def\He#1{{}^{#1}\mbox{He}}

\def\Sunit{\mbox{$10^{-20}$ keV-b}}
\def\vsigma{{\mbox{\boldmath $\sigma$}}}
\def\CPT{{\small $\chi$PT}}

\begin{document}

\input epsf.tex    

\input psfig.sty

\jname{Annu. Rev. Nucl. Part. Sci.}
\jyear{2004}
\jvol{1}
\ARinfo{1056-8700/97/0610-00}

\title{The Solar $Hep$ Process}

\markboth{Kubodera \& Park}{The Solar $Hep$ Process}

\author{
Kuniharu Kubodera
\affiliation{Department of Physics and Astronomy,
University of South Carolina, Columbia,
SC 29208;
e-mail: kubodera@sc.edu}
Tae-Sun Park
\affiliation{Schhol of Physics,
Korea Institute for Advanced Study,
Seoul 130-012, Korea;\\
e-mail: tspark@kias.re.kr}}


\begin{abstract}
The $Hep$ process is a weak-interaction reaction,
$^3{\rm He}\,+\,p \rightarrow
\,^4{\rm He}+e^++\nu_e$, which occurs in the sun.
There is renewed interest in $Hep$
owing to current experimental efforts to extract
from the observed solar neutrino spectrum
information on non-standard physics
in the neutrino sector.
$Hep$ produces highest-energy solar neutrinos,
although their flux is quite modest.
This implies that the $Hep$ neutrios can
at some level influence the solar neutrino spectrum
near its upper end.
Therefore, a precise interpretation
of the observed solar neutrino spectrum
requires an accurate estimate of the $Hep$ rate.
This is an interesting but challenging task.
We describe the difficulties involved
and how the recent theoretical developments
in nuclear physics have enabled us to
largely overcome these difficulties.
A historical survey of
$Hep$ calculations is followed by 
an overview of the latest developments.
We compare the results obtained
in the conventional nuclear physics approach
and those obtained in a newly developed
effective field theory approach.
We also discuss the current status of the experiments 
relevant to $Hep$.  

\end{abstract}


\maketitle

\vspace{3cm}

\section{INTRODUCTION}

The $Hep$ process,
$ {}^3{\rm He}\,+\,p \rightarrow \,{}^4{\rm He}+e^++\nu_e $,
is one of the thermonuclear reactions that occurs
in the sun. To explain why this specific process is of current
interest, we first briefly describe
the standard solar model, the solar neutrinos and neutrino
oscillations.

The sun generates its energy by converting four protons into
an alpha particle,
$4p\rightarrow {}^4{\rm He}+2e^+ +2\nu_e$,
via chains of thermonuclear reactions caused by
weak, electromagnetic, or strong interactions.
The $pp$-chain, shown in Fig.~1,
represents by far the most important scheme by which
the $4p\rightarrow {}^4{\rm He}$ burning takes place in the sun.
To establish how these reactions actually proceed
in the sun,
one must carry out a detailed simulation
in which the radial profiles of the mass density,
temperature, chemical composition, {\it etc.},
are determined in such a manner that
hydrostatic equilibrium is satisfied and
the empirically known solar properties
come out correctly.
The principal inputs that go into this simulation
are the nuclear reaction rates,
equation of state, elemental abundances,
and radiative opacity.
Over the past four decades a great deal of
effort has been invested in this subject,
and out of this endeavor has emerged
a quantitative model of the sun,
called the standard solar model (SSM)
\cite{bu88,bah89,BP00,tur93,bpg03}.
Among many quantities determined by SSM
are the time rates of the
thermo-nuclear reactions occurring in the sun \cite{BP00};
Figure 1 indicates the predicted branching ratios 
of the various paths involved in the $pp$-chain.
Among the reactions featured in Fig.~1,
five are weak-interaction processes
that emit solar neutrinos,
and SSM predicts the flux $\phi_\nu$
from each source of the solar neutrinos \cite{BP00};
this prediction is shown in Fig.~2.
Studying the solar neutrinos is very important
for two reasons.
First, it gives direct information about the physics
of the solar interior,
since the neutrinos exiting the sun
experience hardly any interactions
with the solar medium
other than refractive effects 
(related to the MSW effect to be discussed later).
This should be contrasted
with the behavior of the photons,
which interact with the solar medium so many times
that, by the time they reach the surface
(after $\sim$40,000 years !),
they do not carry much information
about the solar interior.
Second, the solar neutrinos
can provide valuable information
on the properties of the neutrinos themselves;
the sun is an extremely strong neutrino source
and hence can be highly useful
for neutrino physics.

The first measurement of the solar neutrinos
was done by Davis and his collaborators \cite{dav68},
who used a ${}^{37}{\rm Cl}$ target.
The results indicated that 
the sun indeed emits neutrinos whose flux 
is in approximate agreement with the SSM prediction
\cite{bah64},
which supports the basic idea of the thermonuclear
origin of the solar energy.
At a more quantitative level, however, the measured flux
was significantly lower than predicted by SSM.
This deficit, or ``solar neutrino problem",
was also confirmed
by water Cerenkov counter experiments at
the Kamiokande \cite{fuketal96}
and Super-Kamiokande \cite{fuketal98a},
by gallium-target experiments
\cite{abdetal96,ansetal96},
and by heavy-water Cerenkov counter experiments
at the Sudbury Neutrino Observatory (SNO) \cite{ahmetal01}.
It is to be noted that because of different detection
threshold energies (see Fig.~2),
these experiments are sensitive to different regions
of the solar neutrino spectrum.
If we denote by $R$ the ratio of the observed event rate
for a given solar neutrino detection experiment
to the event rate expected from the SSM prediction,
the current status of the solar neutrino problem
is summarized as follows:
$R = 0.34\pm 0.03$ for the chlorine experiment
\cite{cleetal98};
$R= 0.465\pm 0.015$ for the water Cerenkov counter
experiment \cite{fuketal01};
$R= 0.54\pm 0.03$ for the gallium experiments
\cite{gav03,hametal99,bel03};
$R=0.35\pm 0.02$ \cite{ahmetal02a,ahmetal02b}
and $R=0.32\pm 0.02$ \cite{ahmetal03}
for the heavy-water Cerenkov
counter experiments.
It should be mentioned 
that the errors attached to the above values of $R$
only include experimental errors.
Obviously, the degree  of seriousness of
the solar neutrino problem ($R<1$)
hinges on the accuracy
of the SSM predictions.
The latest discussion of the errors to be assigned
to the SSM predictions \cite{BP00}
finds it extremely unlikely that
the solar neutrino deficit can be attributed
to the uncertainties in SSM.
This conclusion is further corroborated
by highly stringent constraints imposed by
the helioseismological data \cite{bah00a}.

The above discussion presupposes
that the neutrinos created in the sun
travel to the terrestrial detectors
without changing their identity (or flavor).
Let us recall that there are three distinct
neutrinos,   
-- electron neutrinos ($\nu_e$), 
muon neutrinos ($\nu_\mu$),
and tau neutrinos ($\nu_\tau$) ---
and that it is the electron neutrinos that
are produced in the sun.
If there exists a mechanism \cite{pon57,nms62}
that changes electron neutrinos
muon neutrinos or tau neutrinos
before they reach a terrestrial detector
that detects only electron neutrinos,
then there would be an effective deficit
of solar neutrinos.
This transmutation of the neutrino flavor,
called neutrino oscillations,
signals physics that goes beyond
the well-established standard model
of particle physics,
and therefore its experimental verification
is of paramount importance.
Neutrino oscillations can occur either during
the neutrino's propagation in vacuum \cite{pon57,nms62},
or as the neutrinos travel in matter
and experience refractive interactions
with the medium (MSW effect) \cite{MSW}.

Now, neutrinos can be detected
either via charged-current (CC) reactions
or via neutral-current (NC) reactions.
Since a CC reaction involves the change
$\nu_x \rightarrow x$
(where $x$ = e$^-$, $\mu^-$, or $\tau^-$),
it can occur only for the electron neutrino;
the muon and tau-lepton
are too heavy to be created
by solar neutrinos.
Meanwhile, an NC reaction that involves
$\nu_x\rightarrow \nu_x$
occurs with the same amplitude
for any neutrino flavor $x$.
At SNO, the CC reaction
$\nu_e d\rightarrow {\rm e}^- pp$
was used to register the electron neutrino
flux $\phi_{\nu_e}$, whereas the NC reaction
$\nu_x d \rightarrow \nu_x np$
was used to determine the total neutrino flux,
$\phi_{\nu,{\rm T}}\equiv
\phi_{\nu_e}\!+\!\phi_{\nu_\mu}\!+\!\phi_{\nu_\tau}$.
The NC reaction data \cite{ahmetal02a}
showed that $\phi_{\nu,{\rm T}}$
agrees very well with the SSM prediction \cite{BP00},
whereas the CC reaction data \cite{ahmetal01} indicated
$R=0.347\pm 0.029$.
These results have firmly established
flavor transmutations in the solar neutrinos.
(For the evidence obtained from comparison 
of the SNO CC reaction data 
and the Super-Kamiokande data \cite{fuketal01}, 
see Reference \cite{ahmetal01}.)
Independent evidence for neutrino oscillations
is known from the study of
the atmospheric neutrinos
at Super-Kamiokande ~\cite{fuketal98b},
and from the study of the reactor neutrinos
at the KamLAND \cite{eguetal03}.

Neutrino oscillations occur if
a neutrino state produced in a weak-interaction
process (``weak eigenstate") is different from
an eigenstate of the Hamiltonian (mass eigenstate),
and if the mass eigenstates
of different neutrinos are not degenerate.
It is conventional to parametrize
the former aspect in terms of mixing parameters,
and the latter in terms of differences
between the neutrino masses squared
($\delta m^2$'s).
Now that the existence of neutrino oscillations
has been established,
the next important challenge
is to determine the accurate values of
the mixing parameters and $\delta m^2$'s,
quantities that should carry valuable
information on new physics beyond the standard model.
(For a recent survey of this topic, see {\em e.g.},
Reference \cite{fy03}.)
The great importance of this determination
makes it highly desirable to assemble
an over-constraining body of data,
and this is one reason
why detailed studies of the solar $Hep$ process
can be important.

In discussing $Hep$,
it is convenient to use
Bahcall {\it et al.}'s latest treatise on
the SSM \cite{BP00} as a basic reference
(to be called BP00).
According to BP00, the neutrino flux
due to $Hep$ is
$\phi_\nu(hep) = 9.4\times 10^3\ {\mbox{cm}}^{-2} {\mbox{s}}^{-1}$,
which is seven orders of magnitude smaller than
the $pp$-fusion neutrino flux,
and three orders of magnitude smaller than
the $^8$B neutrino flux $\phi_\nu(^8{\rm B})$;
the smallness of $\phi_\nu(hep)$
can also be seen from Fig.~2.
(For the radial distribution
of sites of $Hep$ neutrino generation inside the sun,
see Fig.~6.1 in Reference \cite{bah89}.)
Because of its extremely small branching ratio (see Fig.~1),
$Hep$ in fact does not affect solar model calculations.
So why does it interest us ?
One reason is that $Hep$ is a potential source of
useful information on non-standard physics
in the neutrino sector.
$Hep$ generates neutrinos having maximum energy
$E_\nu^{max}(hep)$ = 18.8 MeV,
which is higher than that of the
$^8$B neutrinos,
$E_\nu^{max}(^8{\rm B})$ = 17 MeV;
thus the $Hep$ neutrinos near the upper end of
their spectrum represent the 
highest-energy solar neutrinos
(see Fig.~2).
Solar neutrino detectors such 
as Super-Kamiokande and SNO
can determine the spectrum of the solar neutrinos
in a region dominated by the $^8$B neutrinos.
Meanwhile, the shape of $\phi_\nu(^8{\rm B})$ is
independent of solar models to an accuracy of 1 part in $10^5$
\cite{bah91}.
Therefore, in the absence of $Hep$ neutrinos,
any deviation of the observed $\phi_\nu$
from $\phi_\nu(^8{\rm B})$ in the higher $E_\nu$ region
reflects the non-standard behavior of neutrinos.
Solar neutrino experiments that approach 
the level of precision needed for studying 
this deviation 
have already been reported from Super-Kamiokande
\cite{fuketal99,fuketal01,fuketal02} (see below).
Apart from the ramifications
for neutrino physics,
the study of $Hep$ neutrinos
is also important as a possible additional check 
of the SSM itself.

In interpreting these and future experiments,
we need to know accurately to what extent 
the $Hep$ neutrinos can
affect $\phi_\nu$ in the ${}^8\mbox{B}$ neutrino region
\cite{kuz65,bu88,efgm98,bk98},
and for this we must make
a reliable estimation of the $Hep$ cross section.
This task, however, turns out to be
extremely challenging.
For one thing,
although the primary $Hep$ amplitude is formally of the
Gamow-Teller (GT) type,
the usually dominant one-body GT amplitude is
strongly suppressed for $Hep$ (see below).
Furthermore, the two-body corrections
to the ``leading" one-body GT term have 
opposite sign,
causing a large cancellation.
It is therefore necessary to
calculate these ``corrections" with great accuracy,
which is a highly non-trivial task.
Thus, from a nuclear-physics point of view,
$Hep$ presents a difficult 
yet very intriguing challenge.

\begin{figure}
\centerline{\psfig{figure=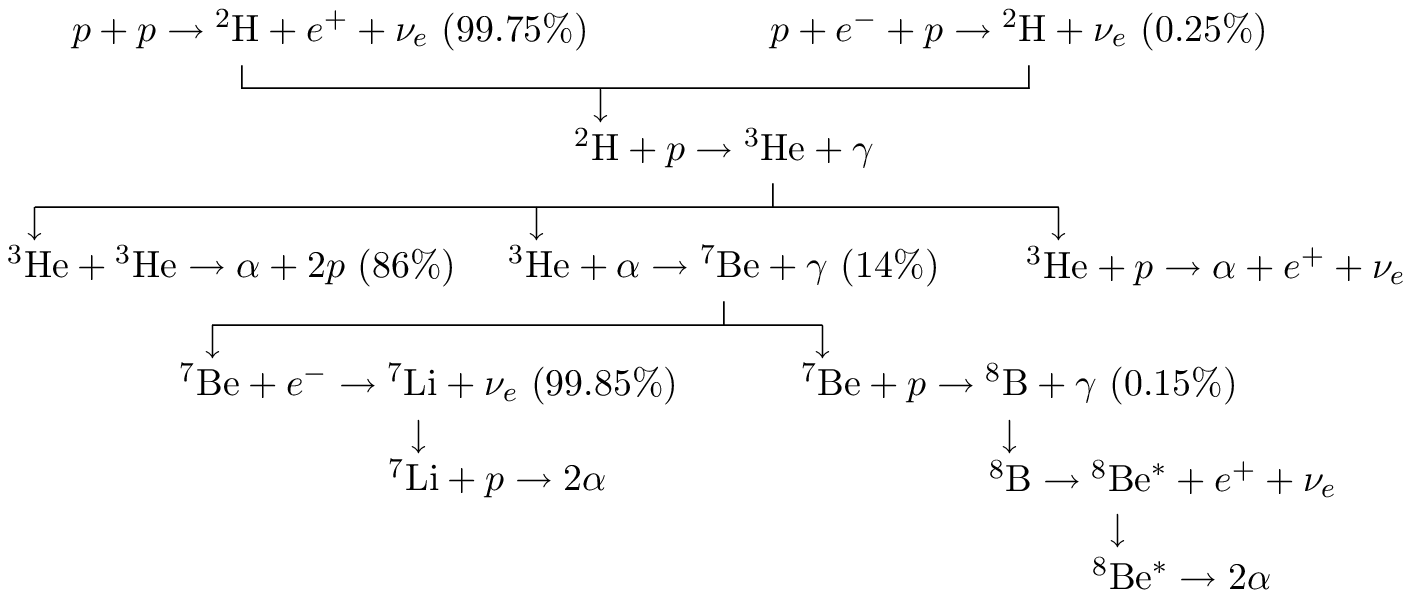,width=4.5in}}
\caption{Solar thermonuclear reactions
in the $pp$-chain and their branching ratios.
The $Hep$ branching ratio is
of the order of 0.01\% or less. }
\label{fig3}
\end{figure}

\begin{figure}
\centerline{\psfig{figure=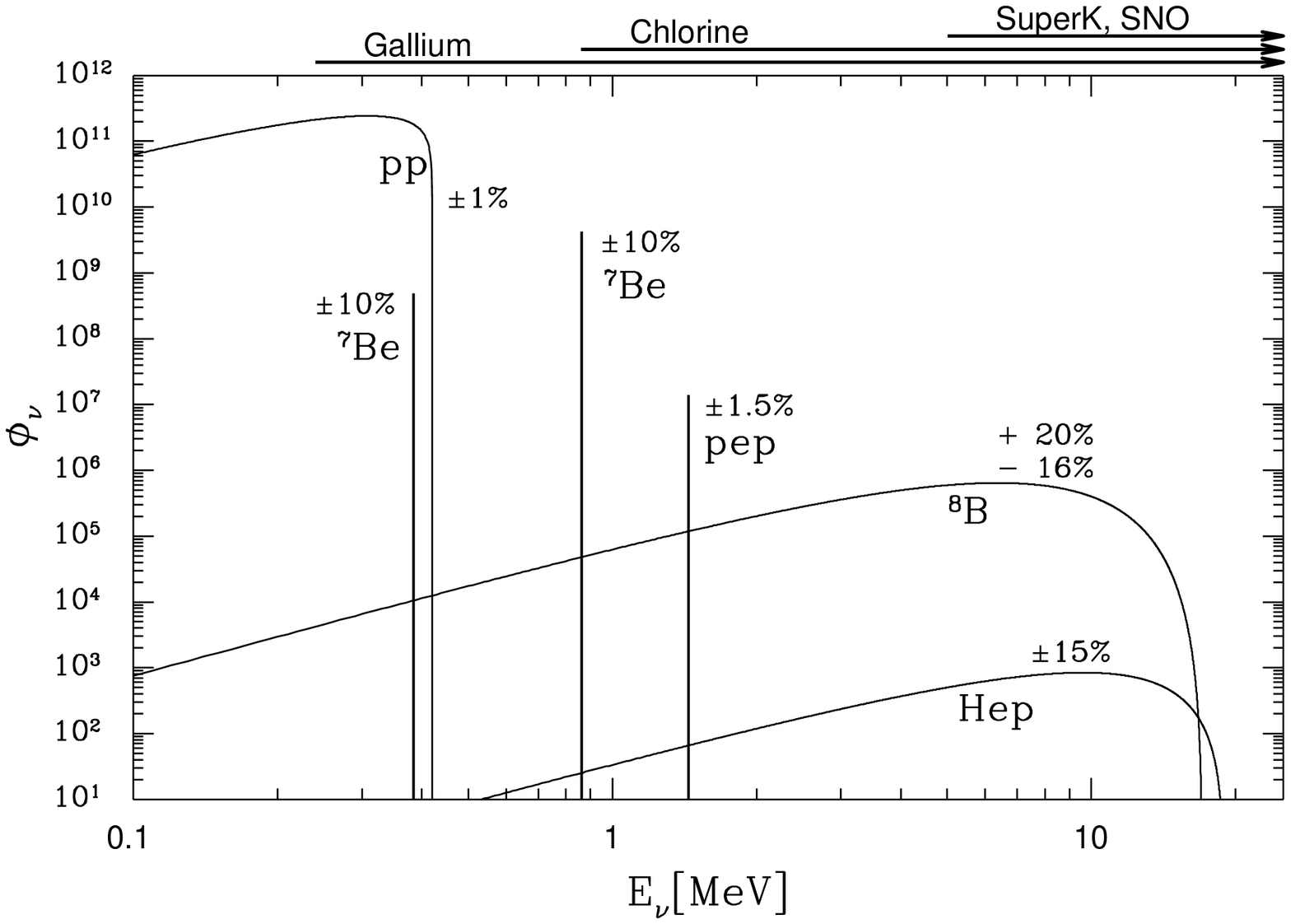,width=4.5in}}
\caption{Solar neutrino spectrum $\phi_\nu$
versus the neutrino energy $E_\nu$.
The neutrino fluxes from continuum sources
are given in units of
${\rm cm}^{-2}{\rm s}^{-1}{\rm MeV}^{-1}$,
and the line fluxes in units of
${\rm cm}^{-2}{\rm s}^{-1}$.
The arrows at the top indicate
the ranges of $E_\nu$ covered by the experiments
mentioned in the text.}
\label{spectrumeps}
\end{figure}

In what follows, we first present a history
of $Hep$ calculations, explaining in more detail
the nature of the difficulties involved
in $Hep$ calculations.
We then describe
how the recent theoretical developments
in nuclear physics have enabled us to
largely overcome these difficulties.
After reporting the latest results obtained
in the so-called standard nuclear physics approach,
we highlight the results
obtained in a newly developed
effective field theory approach.
We then describe the current status of experimental information
on the $Hep$ neutrinos.
At the end we discuss several electroweak processes closely
related to the $Hep$ calculations.

\section{EARLIER CALCULATIONS}

An illuminating survey of the earlier
$Hep$ calculations can be found in
References \cite{bk98,maretal01}.
The $Hep$ reaction rate can be conveniently expressed
in terms of the astrophysical $S$-factor defined by
$S(E)=E\sigma(E)\exp(4\pi\alpha/v_{\rm rel})$,
where $\sigma(E)$ is the $Hep$ cross section at center-of-mass
energy $E$, $v_{\rm rel}$ is the relative velocity
between $p$ and $^3{\rm He}$, and $\alpha$ is
the fine structure constant;
$S(E)$ is a smooth function of $E$
that remains non-vanishing as $E\rightarrow 0$.
The first $Hep$ calculation in 1952
by Salpeter \cite{salpeter}
was based on the extreme single-particle picture
and only considered the overlap
between an $s$-wave proton scattering wave function
and a $1s$ neutron state in $^4$He.
This simplified treatment led to a large value for $S$,
$S(0)= 630\times\Sunit$, and this value was
used by Kuzmin \cite{kuz65},
who was the first to discuss $Hep$
in connection with solar neutrinos.
Werntz and Brennan \cite{wb67}
pointed out the drastic suppression
of the $Hep$ rate due to a specific feature
of the initial and final nuclear wave functions.
The dominant component of $^4$He
has the orbital configuration $(1s)^4$,
which is totally symmetric, {\em i.e.,}
a state with [4] orbital permutation symmetry.
In general, the Pauli principle dictates that
a spin-isospin wave function accompanying
an orbital function with [4] symmetry must be
totally antisymmetric ([1111] state)
and hence must have $S=T=0$.
The contraposition of this property implies that
the $p$-$^3$He state
with isospin $T=1$
cannot have [4] orbital symmetry.
Meanwhile, the one-body GT operator,
$\sum_{i=1}^4\vsigma(i)\tau_-(i)$,
acting only on the spin and isospin,
cannot change the symmetry properties
of orbital wave functions.
For $Hep$, therefore, the leading one-body GT operator
cannot connect the main components
of the initial and final states ---
a feature that leads to a drastic suppression
of the $Hep$ amplitude.
This implies that the
exchange-current (EXC) effects
may play an exceptionally large role here.
Werntz and Brennan \cite{wb67} attempted
to relate the $Hep$ rate to
the M1 matrix element for the $Hen$ process,
where $Hen$ is radiative capture of
a thermal neutron on $\He3$:
$\He3 +n\rightarrow \He4+\gamma$.
They assumed
(a) the validity of isospin symmetry
apart from the difference in the radial functions
of the incident nucleons (proton for $Hep$ and
neutron for $Hen$); and
(b) that two-body EXC terms dominated for both
$Hep$ and $Hen$ and that their matrix elements
could be related to each other via an
isospin rotation.
Based on the upper limit for the $Hen$ cross section
known at that time, Werntz and Brennan
gave an upper limit for the $Hep$ $S$-factor,
$3.7\times \Sunit$,
which was about 200 times smaller
than Salpeter's estimate.
Later Werntz and Brennan \cite{wb73}
refined their estimate
in several respects, including the addition
of the contributions from $p$-wave capture channels,
and they arrived at an $S$-factor of
$8.1\times \Sunit$.
Tegn\'{e}r and Bargholtz \cite{tb83} also attempted
to relate $Hep$ to $Hen$, but they
pointed out the importance of
the contributions due to the $D$-state components
in the $^3$He and $^4$He wave functions,
and they used the EXC operators of the type
derived by Chemtob and Rho \cite{cr71};
Tegn\'{e}r and Bargholtz's estimate was
(4 - 25)$\times \Sunit$,
where the spread corresponded to the range
of the experimental values of the $Hen$ cross section
before 1983.
This result was sharpened by Wolfs {\it et al.}\
\cite{woletal89}, who measured the $Hen$ cross
section precisely and reported a value of
$(54\pm 6)\times 10^{-20}$ $\mu$b;
their estimate of the $Hep$ $S$-factor was
$(15.3\pm 4.7)\times \Sunit$.
Wervelman {\it et al.}\ \cite{weretal91} also made a
precision measurement of the $Hen$ cross section
and obtained
$(55\pm 3)\times 10^{-20}$ $\mu$b,
in good agreement with that obtained
by Wolfs {\it et al.}.
However, Wervelmann {\it et al.} predicted
a $Hep$ $S$-factor of
$(57\pm 8)\times \Sunit$.
These estimates should be considered semi-quantitative, 
since even the estimates
of the one-body terms differ wildly
from model to model,
and furthermore it is known \cite{cr71} that
the EXC for GT transitions
should differ from that
for M1 transitions.
Subsequently, Carlson \etal \cite{CRSW91}
showed that there is a signicant cancellation between
the one-body and two-body terms
and that the use of realistic wave functions
(as opposed to schematic wave functions employed
in the previous calculations)
is crucial for a reliable estimation of the $Hep$ rate.
These authors performed
a variational Monte Carlo calculation
and, with the use of EXC operators
derived from pion- and $\rho$-exchange diagrams
and $\Delta$-excitation diagrams,
they obtained $S= 1.3\times \Sunit$ \cite{CRSW91}.
Schiavilla \etal \cite{SWPC92}
performed a similar calculation
but with the use of explicit $\Delta$ degree of freedom
and obtained
$S =(1.4-3.2)\times\Sunit$.
The $Hep$ calculations up to this point only considered
the contribution of the $s$-wave capture channel,
except in the work of Werntz and Brennan \cite{wb73}.
Horowitz \cite{hor99} presented a new estimate of the
contribution of the $p$-wave capture channel,
using schematic wave functions,
and emphasized that it could be of substantial magnitude.
We categorize $Hep$ estimations
that have appeared since 2000
as ``recent" calculations
and discuss them in the next section.

As the above survey shows, the calculated value
of the $Hep$ $S$-factor changed by
two orders of magnitude
from the original Salpeter value.
Fortunately, however, an encouraging sign of convergence
in the $Hep$ $S$-factor
has been emerging over the past few years.
This is attributable, first,  to further significant progress
along the line of work following 
References \cite{CRSW91,SWPC92}.
Second, the application of effective field theory to $Hep$
has greatly increased the reliability
of the calculated $S$-factor.
These latest developments are the subjects
of the following sections.
To present them coherently,
we first survey the relevant theoretical frameworks
in a somewhat general context, and then proceed to discuss
the specific numerical results for $Hep$.

\section{RECENT CALCULATIONS -- THEORETICAL FRAMEWORK}

\subsection{Standard Nuclear Physics Approach (SNPA)}

The phenomenological potential
picture has been highly successful
in describing a great variety of nuclear phenomena.
In this picture a model Hamiltonian for an $A$-nucleon system
involves a phenomenological
two-body potential $v^{phen}$
(and, if needed, an additional phenomenological
three-body potential).
Once this model Hamiltonian is specified,
the nuclear wave function
is obtained by solving the $A$-body Schr\"{o}dinger equation.
Recent progress in
numerical techniques for this type of calculation
has reached
such a level \cite{cs98} that the wave functions
of low-lying levels for light nuclei
can now be obtained nearly without approximation.
This achievement frees us from the
``usual" nuclear physics complications
that arise from
truncation of nuclear Hilbert space
down to certain model space
(such as shell-model configurations,
cluster-model trial functions, {\it etc.}).
Because there is large freedom in choosing
a possible form for the short-range part of $v^{phen}$,
one assumes a certain functional form
and fixes the parameters appearing in it
by demanding that
the nucleon-nucleon (NN) scattering data and
the deuteron properties be reproduced.
There are by now several so-called high-precision
phenomenological potentials
that can reproduce all the existing two-nucleon data
with normalized $\chi^2$ values close to unity.
In normal circumstances,
nuclear responses
to external electroweak probes are
given, to good approximation,
by one-body terms; these are also called
the impulse approximation terms.
To obtain higher accuracy, however,
one must also consider exchange current (EXC)
terms, which represent the contributions
of nuclear responses involving two or more nucleons.
In particular, if for some reason
the impulse approximation contributions are suppressed,
it becomes essential to take account of the
EXC contributions.
These EXC's are usually derived
from one-boson exchange diagrams,
which impose the low-energy theorems
and current algebra properties
on the vertices featured in the
diagrams \cite{cr71,dor,it79}.
A formalism based on this picture
is referred to as the
standard nuclear physics approach
(SNPA), also called the potential model
in the literature.
SNPA has been used extensively
to describe nuclear electroweak processes
in light nuclei,
and the generally good agreement
between theory and experiment~\cite{cs98}
gives a strong indication
that SNPA captures much
of the physics involved.
The calculations quoted earlier \cite{CRSW91,SWPC92}
represent the early stage of SNPA.

\subsection{Effective Field Theory (EFT)}

Although SNPA has been scoring undeniable successes
in correlating and explaining a vast variety of data,
it is still important from a fundamental point of view
to raise the following issues.
First, since hadronic systems are governed
by quantum chromodynamics (QCD),
one should ultimately be able to relate nuclear phenomena 
to QCD,
but this relation is not visible in SNPA.
In particular, whereas chiral symmetry is known
to be a fundamental symmetry of QCD,
SNPA is largely disjoint from this symmetry.
Second, in SNPA, even for describing low-energy phenomena,
we start with a ``realistic" phenomenological potential
that is tailored to encode short-range (high-momentum)
and long-range (low-momentum) physics
simultaneously.
This mixing of the two different scales
seems theoretically unsatisfactory.
Third, in writing down a phenomenological Lagrangian
for describing the nuclear interaction and
nuclear responses to the electroweak currents,
we find no clear guiding principle in SNPA ---
no obviously identifiable expansion parameter
that helps us to control the possible forms of terms in
the Lagrangian and that provides a general
measure of errors in our calculation.
To address these and related issues,
a new approach based on EFT
was proposed~\cite{wei90},
and it has been studied with great intensity
(for reviews, see References
\cite{bkm}-\cite{br02}).

The intuitive picture of EFT is rather simple.
In describing phenomena
characterized by a typical energy-momentum scale $Q$,
we expect
that we need not include in our Lagrangian
those degrees of freedom
that pertain to energy-momentum scales
much higher than $Q$.
This expectation motivates us to
introduce a cut-off scale $\Lambda$
that is sufficiently large compared with $Q$
and to classify our fields
(to be generically represented by $\Phi$)
into two groups: high-frequency fields
($\Phi_{high}$)
and low-frequency fields
($\Phi_{low}$).
By eliminating (or ``integrating out")
$\Phi_{high}$,
we arrive at an {\it effective} Lagrangian
that only involves
$\Phi_{low}$
as explicit dynamical variables.
Using the notion of path integrals,
we find that the effective Lagrangian ${\cal L}_{{\rm eff}}$
is related to the original Lagrangian ${\cal L}$ as
\bea
\int\![d\Phi]\e^{{\rm i}\int d^4x{\cal L}[\Phi]}
&=&
\int\![d\Phi_{high}][d\Phi_{low}]
\e^{{\rm i}\int d^4x{\cal L}
[\Phi_{high},\Phi_{low}]} \nonumber\\
&\equiv& 
\int\![d\Phi_{low}]
\e^{{\rm i}\int d^4x{\cal L}_{\rm eff}
[\Phi_{low}]}\,.\label{EFTdef}
\eea
One can show that ${\cal L}_{{\rm eff}}$ defined
by eq.(\ref{EFTdef})
inherits the symmetries
(and the patterns of symmetry breaking, if there are any)
of ${\cal L}$.
It also follows that
${\cal L}_{{\rm eff}}$ should be
the sum of all possible monomials of
$\Phi_{low}$ and their derivatives
that are consistent with the symmetry requirements
dictated by ${\cal L}$.
Because a term that involves $n$ derivatives
scales like $(Q/\Lambda)^n$,
the terms in ${\cal L}_{{\rm eff}}$ can be organized
into a perturbative series
in which $Q/\Lambda$ serves as an expansion parameter.
The coefficients of terms in
this expansion scheme are called
the low-energy constants (LECs).
Insofar as  all the LEC's up to a specified order $n$
can be fixed either from theory or from fitting
to the experimental values of the relevant observables,
${\cal L}_{{\rm eff}}$ serves as
a complete (and hence model-independent) Lagrangian
to the given order of expansion.

When EFT is applied to nuclear physics,
the underlying Lagrangian
is the QCD Lagrangian ${\cal L}_{QCD}$,
whereas, for the typical nuclear physics
energy-momentum scale
$Q\ll \Lambda_{\chi}\sim 1$ GeV,
the effective degrees of freedom
that would feature in the effective
Lagrangian ${\cal L}_{{\rm eff}}$
are hadrons rather than the quarks and gluons.
It is not obvious 
how to apply the formal definition, 
eq.(\ref{EFTdef}),
to establsh a relation
between ${\cal L}_{QCD}$ and 
${\cal L}_{{\rm eff}}$ written 
in terms of the hadrons,
since the hadrons cannot be simply
identified with the low-frequency field, 
$\Phi_{low}$, in the original Lagrangian.
At present, the best one can do is
to resort to symmetry considerations
and the above-mentioned expansion scheme. 
Here chiral symmetry plays an important role.  
It is known that chiral symmetry is spontaneously
broken, generating the pions
as Nambu-Goldstone bosons.
This feature can be incorporated
by assigning suitable chiral transformation properties
to the Goldstone bosons
and writing down all possible chiral-invariant
terms up to a specified chiral order 
(see \eg Reference \cite{geo84}).
The above consideration presupposes
exact chiral symmetry in ${\cal L}_{\rm QCD}$.
In reality, ${\cal L}_{\rm QCD}$ contains
small but finite quark mass terms,
which explicitly violate chiral symmetry
and lead to a non-vanishing value
of the pion mass $m_\pi$.
Again, there is a well-defined method
to determine what terms are needed
in the Goldstone boson sector
to represent the effect of
explicit chiral symmetry breaking
\cite{geo84}.
We can then establish a counting rule,
called chiral counting, which allows us
to classify the relative importance of a term
in ${\cal L}_{\rm eff}$ and a given Feynman diagram
according to the number of powers in
$Q/\Lambda$ and $m_\pi/\Lambda$.
These considerations lead to an EFT
called chiral perturbation theory
($\chi$PT)~\cite{wei79,gl84}.

The successes of $\chi$PT in
the meson sector are well known
(see, e.g., Reference \cite{bkm}).
A problem we encounter in extending $\chi$PT
to the nucleon sector is that,
because the nucleon mass $m_N$ is comparable to
the cut-off scale $\Lambda_{\chi}$,
a simple application of expansion in $Q/\Lambda$
does not work.
This problem can be circumvented by employing
heavy-baryon chiral perturbation theory
(HB$\chi$PT) \cite{jm91}.
HB$\chi$PT has been applied with great success
to the one-nucleon sector 
(see \eg Reference \cite{bkm}).
HB$\chi$PT, however, cannot be applied
in a straightforward manner to nuclear systems,
because nuclei involve
very low-lying excited states,
and the existence
of this small energy scale
spoils the original counting rule~\cite{wei90}.

\subsection{Nuclear EFT in the Weinberg Scheme
-- $\Lambda$-Counting}

Weinberg proposed to avoid this difficulty
by classifying Feynman diagrams into two groups,
irreducible and reducible diagrams~\cite{wei90}.
Irreducible diagrams are those
in which every intermediate state
has at least one meson in flight;
all others are classified as reducible diagrams.
We then apply the above-mentioned chiral counting rules
only to irreducible diagrams.
The contribution of all the two-body irreducible diagrams
(up to a specified chiral order)
is treated as an effective potential
acting on nuclear wave functions.
Meanwhile, the contributions of reducible diagrams
can be incorporated
by solving the Schr\"odinger equation.
This two-step procedure may be referred to as
nuclear \CPT\  or, to be more specific,
nuclear \CPT\   in the Weinberg scheme.

To apply nuclear \CPT\ to a process
that involves (an) external current(s),
we derive a nuclear transition operator
${\cal T}$
by evaluating the complete set of
irreducible diagrams
(up to a given chiral order $\nu$)
that involve the relevant external current(s).
For consistency in chiral counting,
the  nuclear matrix element of ${\cal T}$
must be calculated with the use of nuclear
wave functions that are governed
by nuclear interactions that represent
all the irreducible $A$-nucleon diagrams
up to $\nu$-th order.
If this program could be carried out exactly,
it would constitute an {\it ab initio} calculation.
The unambiguous classification of
transition operators according to their chiral orders
is a great advantage of EFT,
which is missing in SNPA.

\subsection{Nuclear EFT in the KSW Scheme
-- $Q$-Counting}

An alternative form of nuclear EFT is
based on the power divergence subtraction (PDS) scheme.
The PDS scheme proposed
by Kaplan, Savage and Wise (KSW)
in their seminal papers~\cite{ksw}
uses a counting scheme (often called $Q$-counting)
that differs from the Weinberg scheme.
Although a great number of important investigations
have used the PDS scheme
(for a review, see \eg Reference \cite{seattle}), here
we are primarily concerned with the Weinberg scheme.
The reason is that the PDS scheme has so far been used
chiefly only for the two-nucleon systems
(see, however, References \cite{kol,bvK02,bedetal02}),
and so at present it is less directly connected
with the $Hep$ process than the Weinberg scheme is;
no $Hep$ calculations based on the PDS scheme exist
in the literature at the time of this writing.

\subsection{Hybrid EFT}

The preceding subsections
emphasize the formal merits of nuclear EFT.
In actual calculations, however,
it is still a major challenge
to generate, strictly within the EFT framework,
nuclear wave functions whose accuracy is comparable
to that of SNPA wave functions
(see section~\ref{outlooks}, however).
A pragmatic solution to this problem
is to use wave functions obtained in SNPA;
we refer to this eclectic approach as hybrid EFT
\cite{pmr95,pkmr98a,pkmr98b}.
Since the $NN$ interactions
that generate SNPA wave functions
accurately reproduce the two-nucleon data,
the use of hybrid-EFT
is almost equivalent to using the empirical data themselves
to control the initial and final nuclear wave functions,
insofar as the off-shell problem (see below)
and the contributions of three-body (and higher-body)
interactions are properly addressed.

\subsection{EFT* or MEEFT}

Hybrid EFT can be used for light complex nuclei 
($A= 3,\, 4,\, ...$)
with essentially the same accuracy and ease
as for the $A$=2 system.
We should emphasize in this connection
that, in $A$-nucleon systems ($A\ge 3$),
the contributions of transition operators
involving three or more nucleons
are intrinsically suppressed
according to chiral counting,
and hence, up to a certain chiral order,
a transition operator in an $A$-nucleon system
consists of the same EFT-based
one-body and two-body terms
that are used for the two-nucleon system.

As mentioned above,
the chiral Lagrangian
is definite only when the values of
all the relevant LECs are fixed,
but there may be cases where this condition
cannot be readily met.
Suppose that a two-body EXC operator
under consideration contains an LEC (call it $\kappa$)
that cannot be determined with the use of
$A$=2 data alone.
It is possible that
an observable (call it $\Omega$)
in a $A$-body system ($A\ge 3$)
is sensitive to $\kappa$ and
that the experimental value of $\Omega$
is known with sufficient accuracy.
Then we can determine $\kappa$
by calculating the hybrid-EFT matrix element
that corresponds  to $\Omega$
and adjusting $\kappa$ to reproduce
the empirical value of $\Omega$.
Once $\kappa$ is fixed this way,
we can make predictions for any other
observables for any other nuclear systems
that are controlled by the same transition
operators.
Hybrid EFT used in this manner is referred to
as EFT* or as MEEFT (more effective EFT).

The effective Lagrangian
${\cal L}_{eff}$ is, by construction, valid
only below the specified cutoff scale $\Lambda$.
This basic constraint should
be respected in nuclear EFT calculations as well.
One way to implement this constraint
is to introduce a momentum-cutoff $\Lambda$
for the two-nucleon relative momentum.
The sensitivity
of the results on the choice of $\Lambda$
is expected to serve as a measure
of uncertainties in the calculational framework.

\section{NUMERICAL RESULTS}
\subsection{$Hep$ Calculation Based on SNPA}

Marcucci \etal have recently carried out
a highly elaborate SNPA calculation of
the $Hep$ rate \cite{maretal00,maretal01}.
The treatment of the four-body wave functions
was improved with the use of
the correlated-hyperspherical-harmonics method
\cite{VKR95}.
The strength of the dominant EXC contribution
due to a $\Delta$-excitation diagram was
adjusted to reproduce the experimental value
of $\Gamma_\beta^{\rm tritium}$,
the tritium beta-decay rate:
$\Gamma_\beta^{\rm tritium}({\rm exp})
=12.32\pm 0.03$ yr \cite{sim87}.
This type of empirical normalization,
first introduced in Reference \cite{CRSW91}, 
is expected to reduce significantly the model dependence of
the calculated $Hep$ rate.
Furthermore, the contribution
of the initial $p$-wave channel
was included.
The resulting $Hep$ $S$-factor at 10 keV
(close to the Gamow peak) is
$S = (10.1\pm 0.6)\times\Sunit$;
this is the value used by Bahcall et al. (BP00)
 \cite{BP00}.
The corresponding threshold value is
$S = 9.64\times \Sunit$.

\subsection{$Hep$ Calculation Based on EFT*}

Park \etal \cite{PMS-hep,PMS-pphep} carried out
an EFT* calculation of the $Hep$ rate up to
next-to-next-to-next-to-leading order
in chiral counting.
To this order, there appears in two-body terms
one LEC that at present cannot be
determined from data belonging to 
the $A$=2 systems.
This unknown LEC,
denoted by $\hat{d}_R$ in Reference \cite{pkmr98b},
parametrizes the strength
of a contact-type four-nucleon axial-current coupling.
Park {\it et al.\ } noted that $\hat{d}_R$
also appears as the only unknown parameter
in the calculation of
$\Gamma_{\beta}^{\rm tritium}$.
They determined $\hat{d}_R$
from the experimental value of
$\Gamma_{\beta}^{\rm tritium}$\cite{sim87}.
With the value of $\hat{d}_R$ determined this way,
Park {\em et al.}\  made a parameter-free EFT* calculation
of the $Hep$ rate~\cite{PMS-hep,PMS-pphep}.
The result for the threshold $S$-factor is
$S = (8.6\pm1.3)\times \Sunit$,
where the error spans the range of
the $\Lambda$ dependence.
This EFT* result supports 
the SNPA results in
References~\cite{maretal00,maretal01}.
It is reasonable 
to expect that, 
if the effects of neglected higher-order terms 
and deviations from the framework of EFT* itself
are sizable, they would cause
the significant $\Lambda$ dependence
in the calculated $S$.
This consideration led Park et al. 
\cite{PMS-hep,PMS-pphep}
to adopt the $\Lambda$ dependence 
($\sim$15 \% variation) of $S_{Hep}$
as a measure of uncertainties in their EFT* calculation
of $S_{Hep}$. 
The above-mentioned large cancellation between
the one-body and two-body contributions in $Hep$
amplifies the $\Lambda$ dependence of
$S_{Hep}$ up to $\sim$15 \%, but this dependence
is still reasonably small. 
(Below we discuss the $pp$-fusion reaction, 
where there is no such cancellation
and the $\Lambda$ dependence is found to
be extremely small.)
In Fig.2, the 15 \% errors
assigned to $\phi_{\nu}(hep)$ reflects the uncertainty
in the EFT* estimation 
of $S_{Hep}$ in References~\cite{PMS-hep,PMS-pphep};
this is the first time that $\phi_{\nu}(hep)$
is presented with an error estimate attached.

\subsection{Off-Shell Problem}

The use of hybrid EFT may bring in
a certain degree of model dependence
due to off-shell effects, because
the phenomenological $NN$ interactions
are constrained only by the
on-shell two-nucleon observables.
This off-shell effect, however, is expected
to be small for the reactions under consideration,
since they involve low momentum transfers
and hence are not extremely sensitive to
the short-range behavior of the nuclear wave functions.
One way to quantify this expectation is to compare
a two-nucleon relative wave function
generated by the phenomenological
potential with that generated by an EFT-motivated potential.
Phillips and Cohen~\cite{pc00} made such a comparison
in their analysis of the one-body operators
responsible for electron-deuteron Compton scattering.
They showed that a hybrid EFT should work well up to
momentum transfer 700 MeV.
A similar conclusion is expected to hold
for a two-body operator,
so long as its radial behavior
is duly ``smeared-out" reflecting
a finite momentum cutoff.
Thus, EFT* as applied to low energy
phenomena is expected to be practically
free from the off-shell ambiguities.

Another indication of the stability
of the EFT* results comes from the recently
developed idea of the ``low-momentum nuclear potential".
Let us recall that a ``realistic phenomenological"
potential $v^{phen}$
is determined by fitting
to the two-nucleon data
up to the pion production threshold energy.
So, physically, $v^{phen}$ should 
reside in a momentum regime
below a certain cutoff $\Lambda_c$.
In the conventional treatment, however,
the existence of this cutoff scale is ignored.
Bogner \etal \cite{bogetal01} proposed to construct an
``effective low-momentum" potential,
$V_{low-k}$, by integrating out from $v^{phen}$
the momentum components higher than $\Lambda_c$.
They calculated $V_{low-k}$s
corresponding to a number of 
well-established $NN$ potentials.
Remarkably, the resulting $V_{low-k}$s were found to lead
to identical half-off-shell $T$-matrices 
for all the cases studied,
even though the ways short-range physics is encoded
in these $v^{phen}$s are quite diverse.
This implies that the $V_{low-k}$s are free from
the off-shell ambiguities, and therefore the use of
$V_{low-k}$s is essentially equivalent to employing
an EFT-based $NN$ potential.
The fact  that EFT* calculations by design contain
a momentum-cutoff regulator essentially
ensures that an electroweak transition matrix element
calculated in EFT* is only sensitive
to those half-off-shell $T$-matrices
that are controlled by $V_{low-k}$,
and therefore the EFT* results
reported by Park et al. \cite{PMS-pphep} are expected to be
essentially free from the off-shell ambiguities.

Furthermore, because correlating the observables
in neighboring nuclei (as was done here)
is likely to serve as
an additional renormalization,
the possible effects of higher-chiral-order terms
and/or off-shell ambiguities
can be significantly suppressed
by the use of EFT*

\section{COMPARISON WITH EXPERIMENTAL DATA}

In Super-Kamiokande experiments,
information on the solar neutrino spectrum
$\phi_\nu$ was obtained
by detecting the recoil electron
in the reaction, $\nu+e^-\rightarrow \nu+e^-$,
for $E_{\rm recoil}\ge 6.5$ MeV
\cite{fuketal99},
and for $E_{\rm recoil}\ge 5$ MeV \cite{fuketal01}.
As mentioned, $\phi_\nu$ in this range is
governed by the $^8$B neutrinos mixed with a tiny
number of $hep$ neutrinos.
Reference \cite{fuketal99} presents the data
in 15 bins between 6.5 MeV and 14 MeV and
one higher-energy bin covering 14 - 20 MeV.
The three highest bins showed a larger number
of events than expected from the then most popular neutrino
oscillation parameters.
Bahcall and Krastev \cite{bk98} analyzed 
these data in detail; they considered various
neutrino oscillation scenarios
and treated the $Hep$ $S$-factor,
$S_{Hep}$, as an adjustable parameter.
The philosophy behind this treatment was that,
although the ``1998 standard value" of $S_{Hep}$ adopted
in \cite{adeetal98}
came from an elaborate SNPA calculation \cite{SWPC92},
a first-principle physics argument
was still needed to
exclude the possibility that $S_{Hep}$ might exceed,
e.g., 10 times this value.
Introducing the enhancement factor $\alpha$
defined by
$\alpha \equiv S_{Hep}/(2.3\times \Sunit)$,
where the denominator is the central value of
the 1998 standard $S_{Hep}$,
Bahcall and Krastev reported that,
by allowing $\alpha$ to be larger than 20,
one could significantly improve  global fits
to all the then available solar neutrino data
for every neutrino oscillation scenario studied.
This result triggered renewed theoretical
efforts by Marcucci {\it et al.}~\cite{maretal00,maretal01}
and Park {\it et al.}~\cite{PMS-hep,PMS-pphep}
to determine $S_{Hep}$ with higher accuracy.
An improved SNPA calculation by
Marcucci {\it et al.}~\cite{maretal00,maretal01}
gave a value of $S_{Hep}$
4.4 times larger than the ``1998 standard value";
the value of $\phi_\nu(hep)^{\rm SSM}$
that appears in BP00 \cite{BP00}
is based on Marcucci {\it et al.}'s $S_{Hep}$.
It is to be noted that the authors of \cite{BP00}
avoided giving an estimate
of total uncertainty in $\phi_\nu(hep)^{\rm SSM}$,
citing the unique subtlety involved in the calculation
of $S_{Hep}$.

According to the more recent Super-Kamiokande
results \cite{fuketal01}
covering $E_{\rm recoil}\ge 5$ MeV,
the observed shape of $\phi_\nu$ is consistent
with an undistorted $^8$B neutrino spectrum shape;
a $\chi^2$ fit of the overall spectrum shape
resulted in $\chi^2/$d.o.f. = 19.1/18,
without any $Hep$ neutrino admixture.
This is not very surprising since $\phi_\nu(hep)$
is typically three orders of magnitude smaller than
$\phi_\nu(^8{\rm B})$.
Meanwhile, by assuming that $1.3\pm2.0$ events registered in
the recoil electron energy bin,
$E_{\rm recoil}=18-21$ MeV,
are due to the $Hep$ neutrinos,
the 90 \% confidence level upper limit
of $\phi_\nu(hep)$ was determined to be
$40\times10^3 {\rm cm}^{-2}{\rm s}^{-1}$ \cite{fuketal01}.
This upper limit is 4.3 times the BP00 prediction
for the no neutrino-oscillation assumption;
thus the experimental $\phi_\nu(hep)$
is consistent with the theoretical value.
However, BP00 stressed that
the significance of this agreement
is limited because one cannot assign
an estimate of uncertainty to the theoretical value
of $S_{Hep}$ used in BP00.

This drawback can be greatly mitigated
with the use of the EFT* calculation of $S_{Hep}$
by Park {\it et al.}~\cite{PMS-hep,PMS-pphep},
which gives
$S_{Hep}$ with a well-controlled error estimate
(in the sense explained earlier).
The use of the central value of $S_{Hep}$ 
obtained by Park et al. \cite{PMS-pphep}
would slightly lower $\phi_\nu(hep)^{\rm SSM}$
but would keep its upper end compatible with
$\phi_\nu(hep)^{\rm SSM}$ in BP00 \cite{BP00}.
Thus, the statement that
the upper limit of the experimental $\phi_\nu(hep)$
is $\sim$4 times the central value of the SSM prediction
remains valid.
However, with the use of $S_{Hep}$ obtained
in the EFT* calculation, the SSM prediction is
controlled within $\sim$15 \% precision,
and this fact is expected
to be valuable in future analyses
of experiments concerning $Hep$ neutrinos.
Coraddu et al. \cite{CLMQ03} pointed out that, 
with the precision of $S_{hep}$ achieved 
in Reference \cite{PMS-pphep}, 
we may be able to use the $\phi_\nu(hep)$ data
to study the possible nonstandard behavior
of the solar core plasma.

\section{RELATED TOPICS}

We have so far concentrated on the calculations of $Hep$.
To further clarify the key aspects involved in these
calculations,
we now discuss the related problems of
evaluating the cross sections
for the $pp$ fusion reaction
($p+p\rightarrow d+e^+ + \nu_e$),
the neutrino-deuteron reactions,
and the $Hen$ reaction.

\subsection{The pp Fusion Process}

The latest SNPA estimation of the $pp$ fusion rate
was carried out by Schiavilla \etal \cite{schetal98},
using the EXC operator whose strength was
normalized to fit $\Gamma_\beta^{\rm tritium}$.
Park {\it et al.}~\cite{PMS-pphep,PMS-pp}
performed an EFT* calculation of 
the $pp$ fusion rate,
using exactly the same method
employed for the $Hep$ calculation.
The result is
$S_{pp}=3.94\!\times\!(1\pm0.005)
\times 10^{-25}\,{\rm MeV\, b}$.
This EFT* result supports the value of $S_{pp}$
obtained in SNPA \cite{schetal98}.
It has been found that $S_{pp}$ in the EFT* calculation
varies by only $\sim$0.1 \%
against changes in $\Lambda$,
and this feature can be regarded as typical 
for unsuppressed transitions such as $pp$ fusion.
Thus, to the extent that the $\Lambda$ dependence
serves as a reasonable measure of theoretical uncertainties
(as discussed above),
the EFT* prediction of $S_{pp}$
can be considered highly robust.
The 0.5 \% error in the above $S_{pp}$
is dominated by the uncertainty 
in the experimental value of 
$\Gamma_{\beta}^{\rm tritium}$.

The PDS scheme also was used to estimate
the $pp$ fusion rate \cite{kr99}.
Taking advantage of very low energy and momentum
involved in this reaction,
Kong and Ravendal used 
``nucleon-only" EFT (without pions).
To the order they considered,
there appears only a single LEC, denoted by $L_{1A}$,
and there have been attempts to constrain
its value using observables
in the two-nucleon systems \cite{L1A}.

\subsection{Neutrino-deuteron reactions}

The $\nu$-$d$ cross sections for $E_\nu\lsim$ 20 MeV
are very important in connection with the SNO experiments.
Nakamura {\it et al.} performed a detailed 
SNPA calculation of the $\nu$-$d$ cross sections 
$\sigma(\nu d)$ ~\cite{NSGK,NETAL},
and Butler et al.~\cite{bc2000}
carried out an EFT calculation of  $\sigma(\nu d)$,
using the PDS scheme~\cite{ksw}.
The EFT results \cite{bc2000}
agree with the SNPA results \cite{NETAL},
if the above-mentioned unknown LEC, $L_{\rm 1A}$,
involved in the PDS scheme
is suitably adjusted.
The optimal value,
$L_{\rm 1A}=5.6\,{\rm fm}^3$, found
by Butler et al. ~\cite{bc2000} is consistent
with the order of magnitude of $L_{\rm 1A}$
expected from the naturalness argument
(based on a dimensional analysis),
$|L_{\rm 1A}|\le 6\,{\rm fm}^3$.
Even though it is reassuring
that $\sigma(\nu d)$ calculated in SNPA and EFT
agree with each other
(in the above-explained sense),
it is desirable to carry out an EFT calculation
that is free from any adjustable LEC.
EFT* allows us
to carry out an EFT-controlled parameter-free calculation
of $\sigma(\nu d)$,
and Ando {\it et al.}~\cite{andetal-nud}
performed such a calculation.
The $\sigma(\nu d)$s they obtained~\cite{andetal-nud}
are found to agree within 1\% with
$\sigma(\nu d)$'s obtained in SNPA~\cite{NETAL}.
Although it is in principle possible to calculate
the tritium beta decay rate in the PDS scheme and 
use $\Gamma_\beta^{\rm tritium}(exp)$
to determine $L_{1A}$,
this program has yet to be carried out.

\subsection{$Hen$ process}
In order to gauge the validity of
a calculational method used for $Hep$,
it is extremely useful to study
as a test case the $Hen$ process,
$\He3+n\rightarrow \He4 +\gamma$.
This is because $Hep$ and $Hen$ involve similar kinematics
and share the characteristic
that the contribution of the normally dominant
one-body transition operator is highly suppressed
owing to the symmetry properties of the wave functions.
An EFT* calculation of $Hen$ has been carried out
by Song and Park \cite{sp03},
and the calculated cross section,
$\sigma=(60.1 \pm3.2\pm1.0)\,\mu{\rm b}$,
is in reasonable agreement with the experimental values,
$(54\pm 6)\, \mu{\rm b}$ \cite{woletal89},
and
$(55\pm 3)\, \mu{\rm b}$ \cite{weretal91}.
This agreement supports 
the validity of the EFT* approach in general
and the EFT* calculation of $Hep$ in particular,
even though there is room for improvements
in the $p$\,-$^3$He 
continuum wave function used in Reference \cite{sp03}.
For the earlier $Hen$ calculations based on SNPA,
see References \cite{CRSW90,SWPC92}.
It is hoped that there will be 
further investigations of $Hen$ 
in both SNPA and EFT*.

\section{SUMMARY AND OUTLOOKS\label{outlooks}}

As exemplified above,
low-energy electroweak processes
in light nuclei play important roles
in astrophysics,
and a recently developed EFT-based formalism,
called EFT* or MEEFT (more effective EFT),
can be used profitably to calculate the cross sections
of these processes with high precision.
We have discussed here $Hep$ and a few closely
related reactions, but this new method is expected
to prove useful for other low-energy 
electroweak processes as well.
The numerical results obtained in EFT*
generally support those obtained in the conventional SNPA,
if the strength of the two-body current
is controlled by the empirical value
of an appropriate observable.
It is to be stressed
that EFT* allows us
to make systematic error estimation
of the calculated cross sections,
a feature that is not readily obtainable in SNPA.

From a formal point of view,
one could hope to improve EFT*
by employing nuclear wave functions
determined in an EFT formalism itself
instead of phenomenological wave functions
obtained in SNPA.
In regard to observables that 
do not involve external currents, 
there has been great progress
in building a formally consistent EFT approach
applicable to complex nuclei
\cite{epeetal00,bedetal02}.
It will be highly informative
to apply this type of formalism
to electroweak processes and compare
the results with those of EFT*.

\vspace*{0.5cm}
\noindent
{\bf Acknowledgment}\\
The authors gratefully acknowledge useful
communications with  J.N. Bahcall, M. Rho,
M. Fukugita, R. Schiavilla,
Y. Suzuki, P. Krastev, F. Myhrer, T. Sato, 
V. Gudkov, D.-P. Min, S. Ando,
S. Nakamura and Y.-H. Song.
KK's work is supported in part 
by National Science Foundation
Grant No. PHY-0140214.

\end{document}